\documentstyle[art12,fleqn]{article}
\begin{document}
\vspace*{2.0cm}
\begin{center}
\centering{\Large \bf {Threshold Analysis for the Inverse\\
ac Josephson Effect}}
\end{center}
\vspace*{1.4cm}
\begin{center}
\centering{G. Filatrella$^{a,b}$, B. A. Malomed$^{c}$, and R. D.
Parmentier$^{a}$}
\end{center}
\vspace*{1.0cm}
$^{a}$Dipartimento di Fisica, Universit\`{a} di Salerno, I-84081 Baronissi
(SA), Italy.\\
$^{b}$Physikalisches Institut, Lehrstuhl Experimentalphysik II,\\
\hspace*{0.2cm} Universit\"{a}t T\"{u}bingen, W-7400 T\"{u}bingen,
Germany.\\
$^{c}$Department of Applied Mathematics, School of Mathematical Sciences,\\
\hspace*{0.2cm} Tel Aviv University, Ramat Aviv 69978, Israel.\\
\vspace*{1.0cm}
\section*{Abstract}
The inverse ac Josephson effect involves rf-induced (Shapiro) steps that
cross over the zero-current axis; the phenomenon is of interest in voltage
standard applications. The standard analysis of the step height in current,
which yields the well-known Bessel-function dependence on an effective ac
drive amplitude, is valid only when the drive frequency is large compared
with the junction plasma frequency or the drive amplitude is large compared
with the zero-voltage Josephson current. Using a first-order
Krylov-Bogoliubov power-balance approach we derive an expression for the
threshold value of the drive amplitude for zero-crossing steps that is not
limited to the large frequency or large amplitude region. Comparison with
numerical solutions of the RSJ differential equation shows excellent
agreement for both fundamental and subharmonic steps. The power-balance
value for the threshold converges to the Bessel-function value in the
high-frequency limit.

\vspace{1.0cm}
\begin{center}
\centering{\today}
\end{center}
\newpage

The term `inverse ac Josephson effect' was coined by Levinsen {\em et al.}
\cite{Levinsen77} to describe Shapiro steps \cite{Shapiro63} in the
current-voltage characteristic of a small Josephson junction that cross
over the zero-current axis. The phenomenon was recognized as being
potentially important for voltage-standard applications inasmuch as the
elimination of the dc bias current would eliminate one possible pathway for
the entry of noise into the system. It was studied
extensively---experimentally, analytically, and computationally---by a
number of authors (see, {\em e.g.}, Kautz \cite{Kautz81} plus references
therein). The `standard' analysis that emerged from these studies expressed
the height in current of phase-locked Shapiro steps in terms of Bessel
functions of an effective drive amplitude; this analysis was shown to be
appropriate when the frequency of the ac drive is large compared with the
junction plasma frequency or when the amplitude of the ac drive is large
compared with the zero-voltage Josephson current. Our analysis, instead, is
based on the power-balance formalism developed by Krylov and Bogoliubov
\cite{Scott70}; it is not limited to the large frequency or the large
amplitude regions. Comparison of the predictions of the power-balance
approach with results obtained by direct numerical integration of the model
differential equation typically show agreement to at least three
significant digits. In the high-frequency limit they approach those
predicted by the Bessel-function expressions.

The starting point of our analysis is the usual shunted-junction model of
the small Josephson junction, subjected to a dc-bias current and an
ac-driving current; the corresponding differential equation, in normalized
form, is
\begin{equation} 
\ddot{\phi}+\alpha\dot{\phi}+\sin\phi=\Gamma+\gamma\cos(\omega t) \, ,
\end{equation}
where $\Gamma$ and $\gamma$ are, respectively, the dc- and ac-driving
currents.

The corresponding unperturbed equation, {\em i.e.}, with $\alpha$, $\Gamma$,
and $\gamma$ set to zero, has two types of solutions, {\em viz.},
oscillatory and rotary. Since we are interested in non-zero-voltage states
of the junction we focus on the latter, which have the form
\begin{equation}  
\phi (t)=2\,\rm{am}\,(\frac{t-t_{o}}{k};k) \, ,
\end{equation}
where am() is the Jacobian elliptic amplitude function \cite{B&F} of
modulus $k$, and $t_{o}$ is an arbitrary constant. The instantaneous
junction voltage in this case is thus given by
\begin{equation}  
\dot{\phi}(t)=\frac{2}{k}\,\rm{dn}\,(\frac{t-t_{o}}{k};k) \, .
\end{equation}

We now return to the original model given by Eq. (1). An obvious physical
requirement for the junction voltage to be a stationary, periodic function
similar to that given by Eq. (3) is that the average power dissipated, {\em
i.e.}, the conductance times the mean-square voltage, be equal to the
average power furnished by the drive currents, {\em i.e.}, the mean-value
of the drive current times the voltage. Assuming that Eq. (2) can be used
as a first approximation to the solution of Eq. (1) in the presence of
dissipation and drive, we can write, using results from \cite{B&F}, the
time-average power dissipated as
\begin{equation}  
P_{out}=\alpha <\!\dot{\phi}^{2}\!>\,=\frac{4\alpha E(k)}{k^{2}K(k)} \, ,
\end{equation}
where $K(k)$ and $E(k)$ are, respectively, the complete elliptic integrals
of first and second kinds. Assuming first the presence of only a dc-bias
current, {\em i.e.}, setting $\gamma=0$, we can write the time-average
input power as
\begin{equation}  
P_{in,dc}=\Gamma <\!\dot{\phi}\!>\,=\frac{\pi \Gamma}{kK(k)} \, .
\end{equation}
Equating Eq. (4) to Eq. (5) gives an expression for the McCumber branch of
the current-voltage characteristic of the junction in the following
parametric form
\begin{equation}  
\Gamma=\frac{4\alpha E(k)}{\pi k} \, ,
\end{equation}
\begin{equation}  
V \equiv \, <\!\dot{\phi}\!>\,=\frac{\pi}{kK(k)} \, .
\end{equation}
In the zero-bias configuration, instead, {\em i.e.}, with $\Gamma=0$ and
$\gamma \neq 0$, we can write the input power as
\begin{equation}  
P_{in,ac}=\gamma <\!\dot{\phi}\,\cos (\omega t)\!> \, ,
\end{equation}
which, by expressing the Jacobian dn() function in terms of its
Fourier-series expansion \cite{B&F}, we can write as
\begin{equation}  
P_{in,ac}=\,<\!\frac{\pi \gamma \cos (\omega t)}{kK(k)}+ \frac{4\pi \gamma
\cos (\omega t)}{kK(k)}\sum_{m=1}^{\infty}\frac{q^{m}}{1+q^{2m}} \cos
[\frac{m\pi (t-t_{o})}{kK(k)}]\!> \, ,
\end{equation}
where $q \equiv \exp (-\pi K(k')/K(k))$, in which $k'$ is the complementary
modulus \cite{B&F}. From Eq. (9) we see that $P_{in,ac} \neq 0$ only if
\begin{equation}  
\omega = \frac{m\pi}{kK(k)}
\end{equation}
for some integer $m$, and, assuming Eq. (10) to be satisfied, that
$P_{in,ac}$ can vary smoothly from zero to $(P_{in,ac})_{max}$ depending on
the value of the phase-shift term $t_{o}$ (we have $P_{in,ac}= (P_{in,ac})
_{max}$ when $t_{o}=0$). The threshold value, $\gamma_{thr}$, of the drive
amplitude is the minimum value for which it is possible to satisfy the
equation $P_{out}=(P_{in,ac})_{max}$; this yields the value
\begin{equation}  
\gamma_{thr}=\frac{2\alpha E(k)}{\pi k}\,\frac{1+q^{2m}}{q^{m}} \, .
\end{equation}
For $\gamma \geq \gamma_{thr}$, the constant $t_{o}$ adjusts itself
according to the relation
\begin{equation}  
\cos [\frac{m \pi t_{o}}{kK(k)}]=\frac{\gamma_{thr}}{\gamma} \, .
\end{equation}

Finally, in the general case in which both $\Gamma \neq 0$ and $\gamma \neq
0$, the dissipation defined by Eq. (4) is balanced by both dc and ac
inputs. This permits, {\em e.g.}, calculating the minimum bias-current
value to which a phase-locked step extends: assuming Eq. (10) to be
satisfied, we find
\begin{equation}  
\Gamma_{min}=\frac{4\alpha E(k)}{\pi k}-\frac{2\gamma q^{m}}{1+q^{2m}} \, ,
\end{equation}
which, for $\gamma > \gamma_{thr}$ [see Eq. (11)], is a negative number.

The `standard' Bessel function expression for the height in current of a
phase-locked Shapiro step is \cite{Kautz81,Braiman80}
\begin{equation}  
\Delta \Gamma=2J_{n}(\frac{\gamma}{\omega \sqrt{\omega^{2}+\alpha^{2}}}) \,,
\end{equation}
where $J_{n}$() is the Bessel function of order $n$. Thus, from Eqs. (6) and
(13), the threshold value of the drive amplitude in this approach is
\begin{equation}  
J_{n}(\frac{\gamma_{thr, B}}{\omega \sqrt{\omega^{2}+\alpha^{2}}}) =
\frac{4\alpha E(k)}{\pi k} \, .
\end{equation}

At this point it must be mentioned that the integer $n$ in Eq. (14) is {\em
not} the same as the integer $m$ in Eq. (10): in fact, the Bessel function
index $n$ refers to a {\em super\/}harmonic number whereas the power
balance index $m$ refers to a {\em sub\/}harmonic number. To be perfectly
clear on this point, the corresponding Shapiro steps in the current-voltage
characteristic of the junction occur at normalized voltages
\begin{equation} 
V=<\!\dot{\phi}\!>=\frac{n}{m}\omega \, .
\end{equation}
Thus, the only place where the two expressions, Eq. (11) and Eq. (15), can
be compared directly is at the fundamental frequency, where $n=m=1$.

Our results are summarized in Fig. 1a,b. In these figures, the solid curve
is the power-balance threshold, calculated from Eq. (11), the dashed curve
is the Bessel-function threshold, calculated from Eq. (15), and the
diamonds are values obtained from the direct numerical integration of Eq.
(1). Except for the highest point in Fig. 1b, where the numerical result
begins to diverge slightly from the power-balance prediction, these two
typically agree to at least three significant digits. A similar agreement
between power-balance prediction and numerical experiment is obtained at
the second subharmonic ($m=2$). For higher frequencies than those shown in
Fig. 1a, the two predicted threshold values, Eq. (11) and Eq. (15),
converge to a common asymptotic limit.

In order to calculate numerically the points indicated by diamonds in Fig.
1, we often found it necessary to `tune' fairly precisely the initial
conditions used for Eq. (1). This fact suggests that, at least in some
cases, the basin of attraction of the phase-locked state is rather
small, perhaps indeed vanishing, a fact that would have important
implications for the stability of the locked state, {\em e.g.}, against
thermal fluctuations. The technique of cell mapping \cite{Soerensen88}
might be a useful tool for exploring this question.

Braiman {\em et al.} \cite{Braiman80} have extended the Bessel function
approach to subharmonic frequencies. In fact, their Eq. (2) has just the
form of the Fourier-series expansion of the Jacobian am() function of our
Eq. (2) \cite{B&F}. This approach gives expressions for the step height in
current---corresponding to Eq. (14) for superharmonic steps---in terms of
sums of products of Bessel functions. The approach, however, clearly becomes
rather unwieldy if more than the first few terms in the Fourier expansion
are employed.

The power-balance approach can also be extended beyond the lowest level of
approximation that we have employed here \cite{Minorsky}; this would
presumably permit a description of superharmonic steps, which presently is
lacking from our analysis. However, superharmonic steps occur at
progressively higher voltages, which implies progressively higher
frequencies, where the simple Bessel-function expression, Eq. (15), is
known to give a reasonable description of the situation. Thus, the reward
to be obtained here might not be worth the effort.

Finally, we must offer one {\em caveat\/}: as is apparent from Fig. 1a, the
power-balance prediction becomes progressively better than the
Bessel-function prediction as the ac drive frequency is reduced below unity
(which is the plasma frequency, with our normalization). This, however, is a
region where, in addition to simple periodic solutions of Eq. (1),
exemplified by the ansatz of Eq. (2), there are known to exist also
complicated quasi-periodic and chaotic trajectories \cite{Kautz85}.
Consequently, whereas Eq. (11) does give a good estimate of the threshold
value of the drive amplitude {\em if} a simple step exists, it does {\em
not} guarantee the existence of such a step.

\vspace*{0.3cm}

We wish to thank Roberto Monaco for a critical reading of the manuscript
and Niels Gr\o nbech-Jensen for illuminating discussions and technical
assistance. B.A.M. thanks the Physics Department of the University of
Salerno for hospitality during the visit that originated this work.
Financial support from the EC under contract no. SC1-CT91-0760 (TSTS) of
the ``Science'' program and from the ``Human Capital and Mobility" program,
from MURST (Italy), and from the Progetto Fianalizzato ``Tecnologie
Superconduttive e Criogeniche'' del CNR (Italy) is gratefully acknowledged.

\newpage
\section*{Figure Captions}
Fig. 1. Comparison of power-balance prediction (solid curve) and
Bessel-function prediction (dashed curve) with numerical integration result
(diamonds), at fundamental frequency ($n=m=1$). Parameters: (a) $\alpha =
0.01$; (b) $\omega = 1.0$.

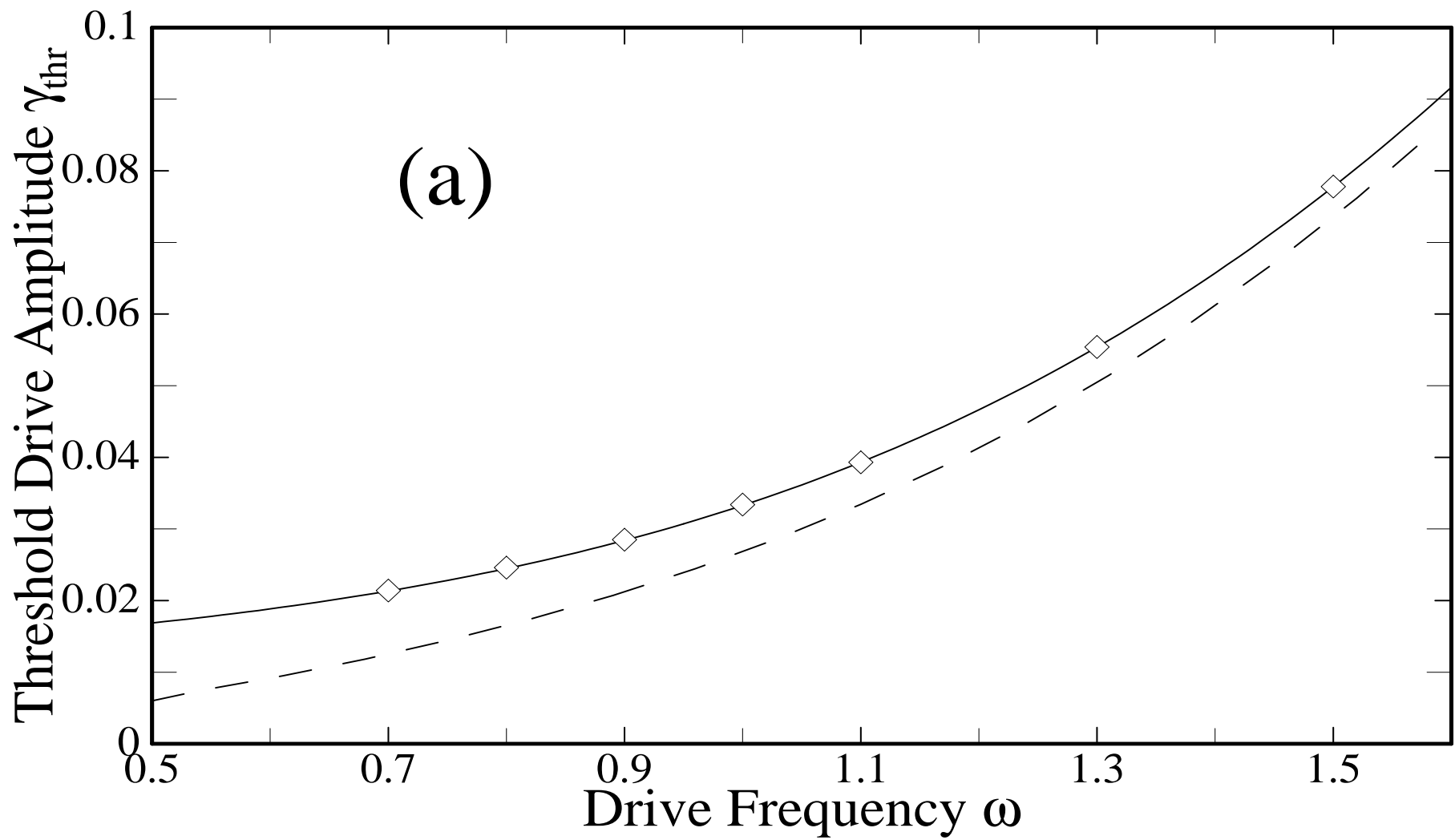
\end{document}